\documentclass[fleqn,usenatbib,letters]{mnras}

\usepackage{newtxtext,newtxmath}
\usepackage[T1]{fontenc}

\DeclareRobustCommand{\VAN}[3]{#2}
\let\VANthebibliography\thebibliography
\def\thebibliography{\DeclareRobustCommand{\VAN}[3]{##3}\VANthebibliography}

\usepackage{graphicx}	% Including figure files
\usepackage{amsmath}	% Advanced maths commands
\newcommand     {\added}[1]   {#1}

\title[The small galactic boxy/peanut bulge]{The small boxy/peanut structure of the Milky Way traced by old stars}

\author[Semczuk et al.]{%
Marcin Semczuk,$^{\!\!1}$
Walter Dehnen,$^{\!\!2,3,1}$
Ralph Sch{\"o}nrich,$^{\!\!4}$ and
E. Athanassoula,$^{\!\!5}$
\smallskip
\\
$^1$ School for Physics and Astronomy, University of Leicester, University Road, LE1 7RH, UK \\
$^2$ Astronomisches Recheninstitut, Zentrum f{\"u}r Astronomie der Universit{\"a}t Heidelberg, M{\"o}nchhofstra\ss{}e 12-14, 69120, Heidelberg, Germany\\
$^3$ Universit{\"a}ts-Sternwarte M{\"u}nchen, Scheinerstra\ss{}e 1, 81679, M{\"u}nchen, Germany\\
$^4$ Mullard Space Science Laboratory, University College London, Holmbury St.~Mary, Dorking, Surrey, RH5 6NT, UK \\
$^{5}$Aix Marseille Université, CNRS, CNES, LAM, Marseille, France
}

\date{Accepted XXX. Received YYY; in original form ZZZ}

\pubyear{2021}

\begin{document}
\label{firstpage}
\pagerange{\pageref{firstpage}--\pageref{lastpage}}
\maketitle

\begin{abstract}
We analyse the positions of RR Lyrae stars in the central region of the Milky Way. In addition to the overall bar shape detected previously, we find evidence for a peanut shaped structure, in form of overdensities near \added{$\ell=-2\degr$ and $1\degr$ at $b\sim-3.5\degr$}. The corresponding physical distance between the two peaks of the peanut is $\sim0.7\,$kpc, significantly shorter than that found from near-IR images (3.3\,kpc) and red-clump stars. Qualitatively this is expected from `fractionation' of bar orbits, which we demonstrate to be matched in a simulation of an inside-out growing disc subsequently forming a bar. 
\end{abstract}


%%%%%%%%%%%%%%%%%%%%%%%%%%%%%%%%%%%%%%%%%%%%%%%%%%

%%%%%%%%%%%%%%%%% BODY OF PAPER %%%%%%%%%%%%%%%%%%

\section{Introduction}
Most bars in \added{Milky-Way size galaxies} have a boxy/peanut (BP) inner bulge \citep{Bureau1999, Erwin2017}, which has about \added{0.4-0.8 times} the \added{length} of the bar \citep{Athanassoula2015}. The Milky Way (MW) is no exception with its peanut structure clearly visible in near-IR images of the bulge region \citep{Weiland1994}. The Galactic peanut has also been dicerned in 3D: \citep{McWilliamZoccali2010, SaitoEtAl2011, Ness2012} linked the observed \citep[][]{Nataf2010} split in the magnitude distribution of red-clump stars to the X-shaped bar/bulge. Furthermore, \citeauthor{Ness2012} noted that this split is clear for stars with $[\mathrm{Fe/H}]>-0.5$, but not for those with $[\mathrm{Fe/H}]<-0.5$. This was the first evidence -- well before the term was introduced by \cite{Debattista2017} -- that `fractionation' has taken place, presumably due to the effect of the bar on the orbits. This was later confirmed in a number of studies, relying on observations and/or simulations. 
    
The results of this fractionation can be witnessed in at least two different, yet clearly related, ways. The simplest for observations is to use metallicity \citep{Babusiaux2016, Ness2012, Uttenthaler2012, Rojas-Arriagada2014}. For simulations, on the other hand, the most convenient and more accurately obtained such property is stellar age \citep{Athanassoula2018, Buck2018}, although some chemodynamic simulations made it possible to base the analysis also on metallicity \citep[e.g.][]{Debattista2017, Athanassoula2017,Fragkoudi2020}.
 
Here, we use RR Lyrae, rather than the more often used RGB stars, and extend previous comparisons of simulations to observations. In Sect. 2, we present our high resolution
simulation. In Sect. 3 we present the data and in
Sect. 4 the BP structure of the various populations. We compare with
observations in Sect. 5 and give a global discussion in Sect. 6.

%%%%%%%%%%%%%%%%%%%%%%%%%%%%
\section{Simulations}
\label{sec:sims}

For comparison with the data, we generated a suite of simulations with a range of disc growing recipes using the $N$-body code \textsc{Griffin}, which uses the fast multipole method as force solver \citep{Dehnen2000} and is tuned to give high force accuracy.

We grow the stellar discs throughout the simulation using the recipe of \cite{AumerSchoenrich2015}. Stars are continuously added mimicking continuous star formation without requiring costly hydrodynamics. Stars are born on near-circular orbits, using the estimated local circular velocity and a velocity dispersion of $8.9$\,km\,s$^{-1}$. Star formation rates are set by equation (3) of \cite{AumerSchoenrich2015} with normalization $16.7\;\mathrm{M_{\odot}}/\mathrm{yr}$ and decay time-scale 8 Gyr. Their distribution is drawn from an inside-out growing exponential disc; its scale length is set by equation (1) of \cite{SchoenrichMcmillan2017}, growing from 0.25\,kpc to 3.75\,kpc in 12.5\,Gyr. To emulate gas depletion and star formation suppression by the bar, no stars are formed at radii $0.1<R/R_\mathrm{CR}<0.7$, where the co-rotation radius $R_\mathrm{CR}$ is inferred from the $N$-body force field (giving $v_\mathrm{circ}(R)$) and the bar pattern speed obtained from the $m=2$ Fourier \added{component of the surface density} once \added{its} amplitude has reached 10\% of the $m=0$ \added{component}.

%%%%%%%%%%%%%
\begin{figure*}
    \centering
    \includegraphics[width=17.85cm]{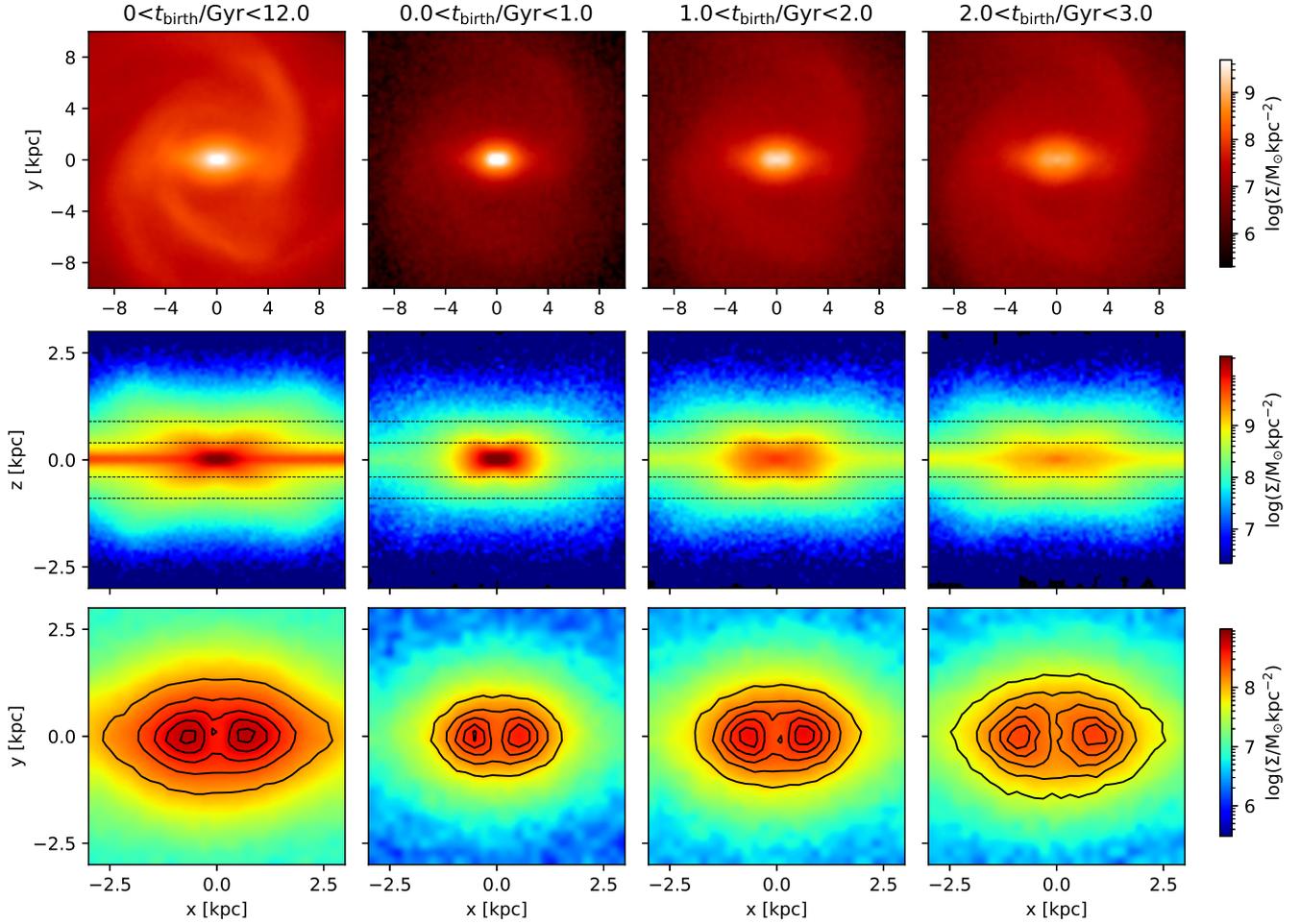}
    \caption{\textbf{Top:} Face-on surface density distributions for the simulated galaxy at $t=12\,$Gyr for four different stellar age bins, with the left panel showing all ages. \textbf{Middle:} Side-on surface density distributions for the same age bins. \textbf{Bottom:} Face-on surface density distributions for the same age bins but restricted to $0.9\,$kpc$>|z|>0.4\,$kpc (indicated by dashed lines in the middle panels) and zoomed in to emphasize how the separation of the boxy/peanut structure changes with stellar age. Units of the colour bars are given for the last 3 columns of the panels, for the leftmost panel the units should be multiplied by two.}
    \label{sim}
\end{figure*}

\begin{figure}
    \centering
    \includegraphics[width=\columnwidth]{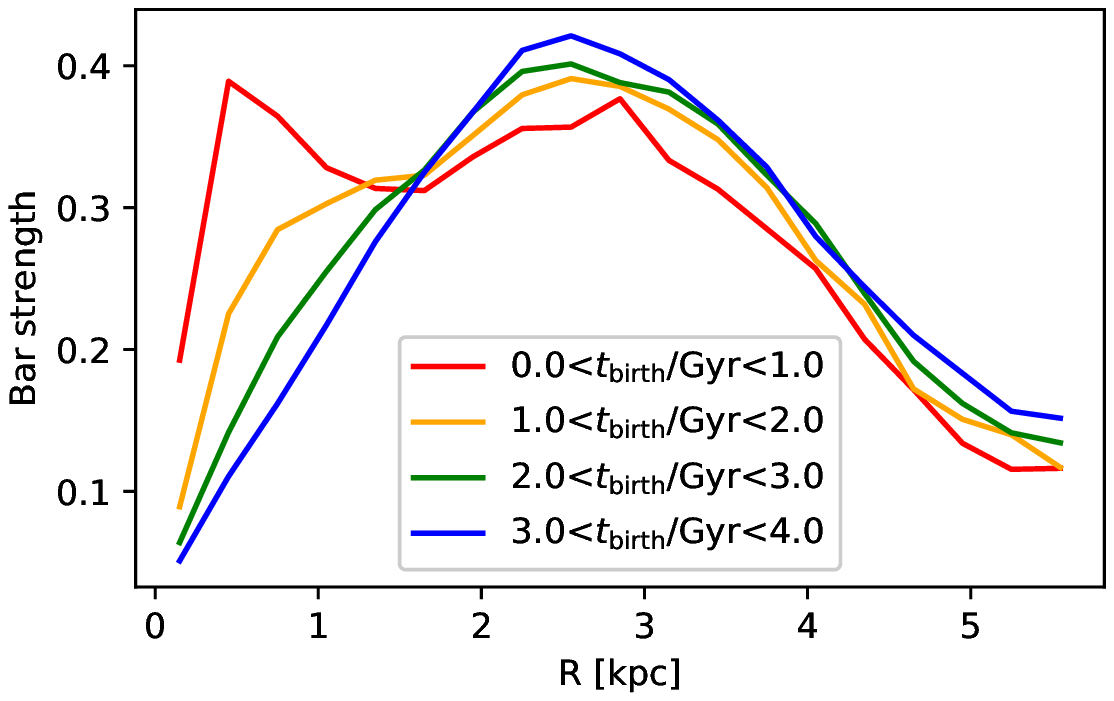}
    \includegraphics[width=\columnwidth]{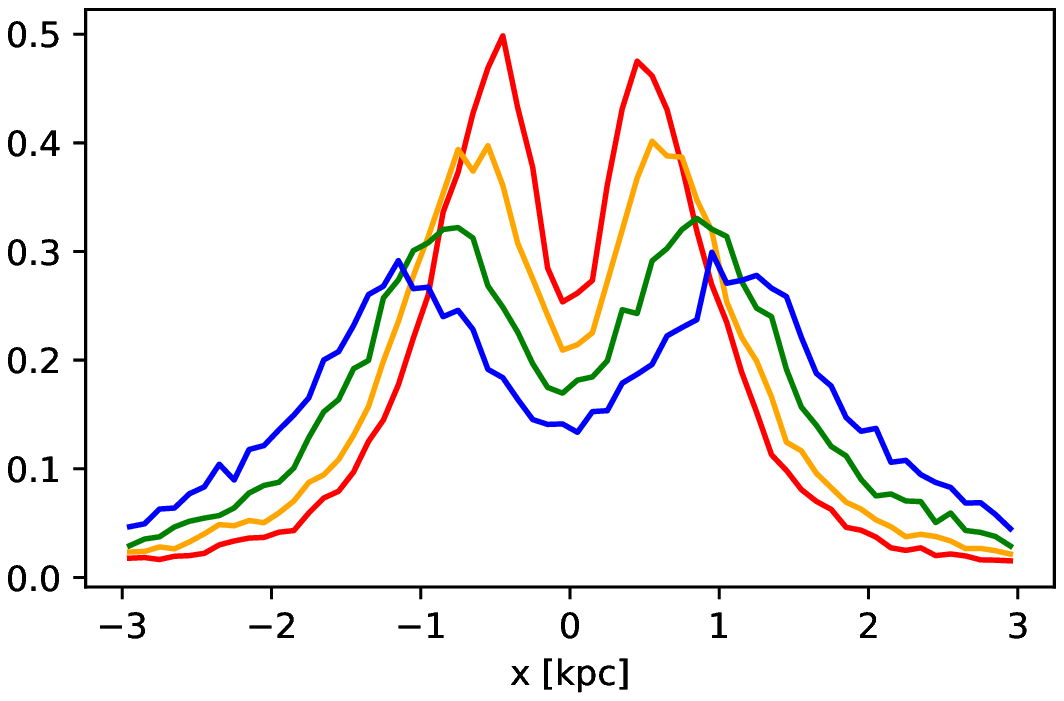}
    \caption{\added{\textbf{Top:} Profiles of relative Fourier amplitude $|\Sigma_2|/|\Sigma_0|$ for four consecutive populations from the simulation. \textbf{Bottom:} Normalized profiles along the bar for the same populations restricted to 0.9 kpc$\, > |z| > 0.4$\,kpc and $|y|$<0.25 kpc (perpendicular to the bar). 
    }}
    \label{fig:bar:profiles}
\end{figure}
%%%%%%%%%%%%%%

All simulations are embedded in a live dark matter halo, which allows the bar to slow and grow by angular momentum transfer. The halo is initialised from a \cite{DehnenMcLaughlin2005} model with scale radius $R_\mathrm{s}=31.25\,$kpc, truncated at $10R_\mathrm{s}$, and circular velocity $v_\mathrm{circ}(R_\mathrm{s}) = 126.6\,\mathrm{km\,s^{-1}}$.
Disc particles have masses $1.2\times10^4\;\mathrm{M_{\odot}}$ and softening length $\epsilon = 0.05 {\rm kpc}$, while the halo particles have masses of $2.7\times10^5\;\mathrm{M_{\odot}}$ and softening lengths of $\epsilon = 0.2{\rm kpc}$. We use 4.2\,M dark matter and (at the end of the simulation) 6.7\,M stellar particles.

\added{
These simulations are computationally much cheaper than hydro-dynamical simulations, but neglect the gas angular-momentum transport. We ran many of them. Here, we present just one exemplary case with parameters chosen such that in 11-12\,Gyr of evolution the bar reaches a length of \added{$\sim4\,$}kpc and $R_\mathrm{CR}$\,=\,6-7\,kpc.}
%%%%%%

%%%%%%%%%%%%%%%%%%%%%%%%%%%%%
\section{Data selection}
\label{sec:sample}
We use the RR Lyrae catalogue \citep{Soszynski2014,Soszynski2019} of the OGLE IV survey \citep{Udalski2015}, more specifically its subset of ab-type RR Lyrae stars (RRab). These same data were previously employed to demonstrate that this population is part of the Galactic bar \citep[based on the ellipsoidal shape and its orientation close to the bar angle,][]{Pietrukowicz2015, Du2020}. 

We derive distances to the RRab stars using prescriptions by \citeauthor{Pietrukowicz2015} and \citeauthor{Du2020}, which take advantage of the  relations for absolute magnitudes by \cite{Catelan2004}, reddening maps by \cite{Gonzalez2012}, and a formula for metallicity by \cite{Smolec2005}. Following \citeauthor{Pietrukowicz2015}, the data are cleaned in the colour-magnitude diagram and globular clusters members rejected.

Finally, we select stars in the latitude range $-6.5\degr<b<-2.8\degr$ to avoid survey incompleteness closer to the Galactic plane. These procedures gave a total sample of 6958 RRab stars. \added{In the analyses below, this number is lowered further by additional restrictions in $\ell$, $b$, or Cartesian coordinates.}

%%%%%%%%%%%%%%%%%%%%%%%
\section{Boxy/Peanuts of different populations}
\label{sec:peanut}
Several studies have demonstrated in $N$-body and hydrodynamical simulations how kinematic fractionation affects the bar and boxy/peanut bulge morphology of different stellar populations \citep{Debattista2017, Fragkoudi2017, Athanassoula2017, Buck2018}. This is no different in our simulations with a growing disc: Fig.~\ref{sim} shows bar and boxy/peanut bulge morphologies for populations depending on time of birth $t_\mathrm{birth}$\added{, while Fig.~\ref{fig:bar:profiles} plots the profiles of the relative $m=2$ Fourier amplitude and the density along the bar major axis for the same populations plus a fourth.} The older stars show a \added{slightly} smaller and more concentrated bar than later generations \added{and a clearly smaller} inner peanut structure.  

The bottom row of Fig.~\ref{sim} shows the peanut structure when limiting the $z$ range similarly to the OGLE field. As this cuts across the X structure, there are two separate density maxima. Their separation increases by a factor $\sim2$ between stars born within the first Gyr of the simulation and those born 2\,Gyr later. All simulations show qualitatively this same picture, while the precise size difference depends on the specific bar evolution (which itself depends on various factors like the disc-to-halo mass ratio, halo angular momentum, star formation history, etc.). We further ran simulations, with and without inside-out disc growth, and found that inside-out disc formation significantly increases the separation difference. An obvious explanation is that the inner part of the bar forms from the old inner disc and that bar growth by capture of younger surrounding disc stars hardly alters the occupation of the inner bar orbits.

%%%%%%%%%%%%%%%%%%%%%
\section{Comparison with observations}
\label{sec:comp}

\begin{figure}
    \centering
    \includegraphics[width=\columnwidth]{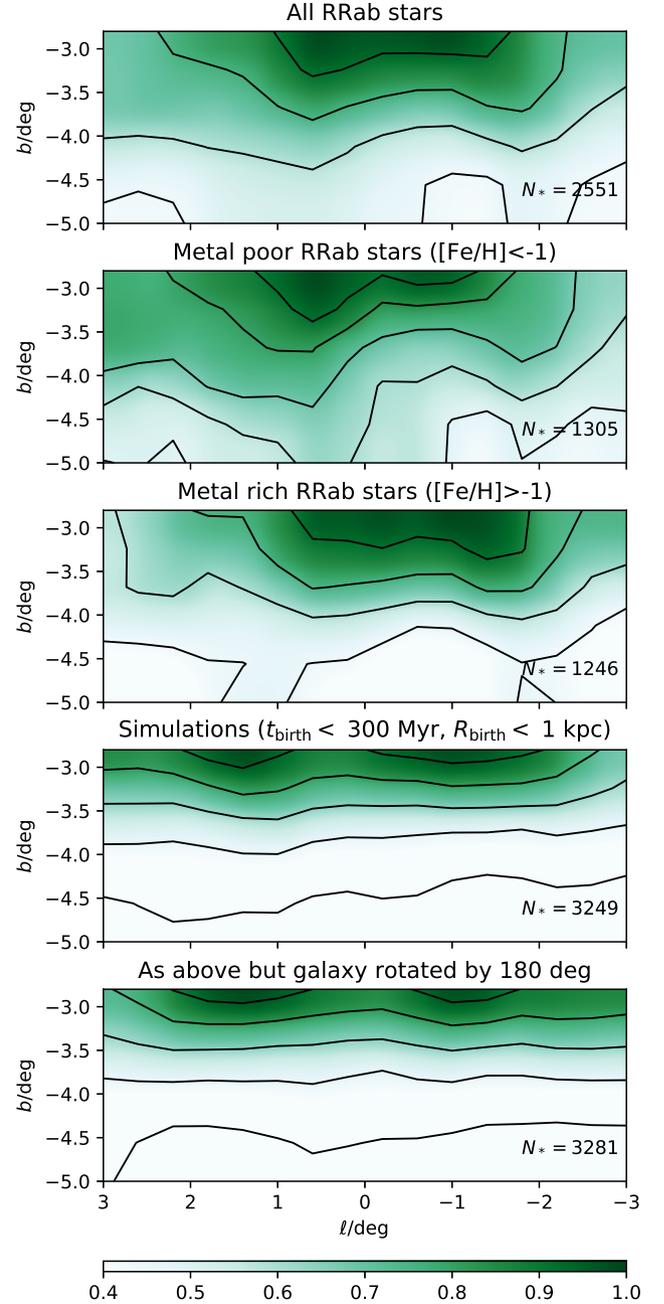}
    \caption{Density in galactic coordinates of different subsets of RRab stars and of two subsets of old star particles from our simulation. The maps were constructed by a kernel density estimator with smoothing length 0.7\degr. Colour scales are linear and normalized to the respective highest value.    }
    \label{front}
\end{figure}

\begin{figure}
    \includegraphics[width=\columnwidth]{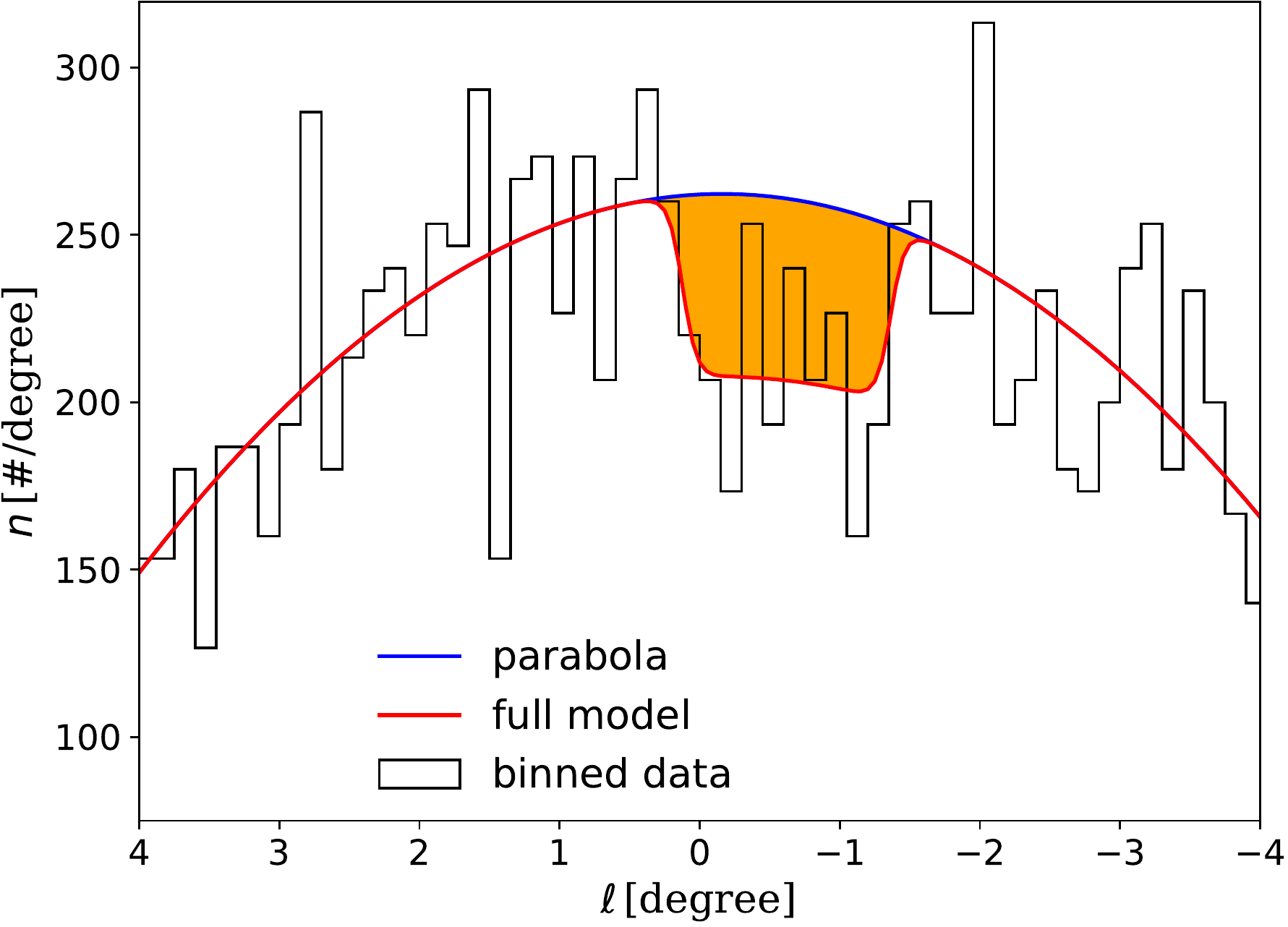}
    \caption{
    \added{
    Distribution over Galactic longitude $\ell$ of OGLE RRab stars at Galactic latitudes $b\in[-5\degr,-3.6\degr]$, and a parametric fit of a parabolic distribution multiplied by a Fermi-distribution shaped trough:
    \[
        h(\ell) = 1 - d + d/[1 + \exp([w-|\ell - \ell_0|]/s)]
    \]
    with parameters $\ell_0$, $w$, and depth $d$, while the steepness is fixed at $s=1/20\degr$. For the best fit $d=0.207\pm0.058$, i.e.\ $d>0$ well above $3\sigma$ significance.}
    }
    \label{fig:minimum}
\end{figure}

\begin{figure*}
    \centering
    \includegraphics[width=18.cm]{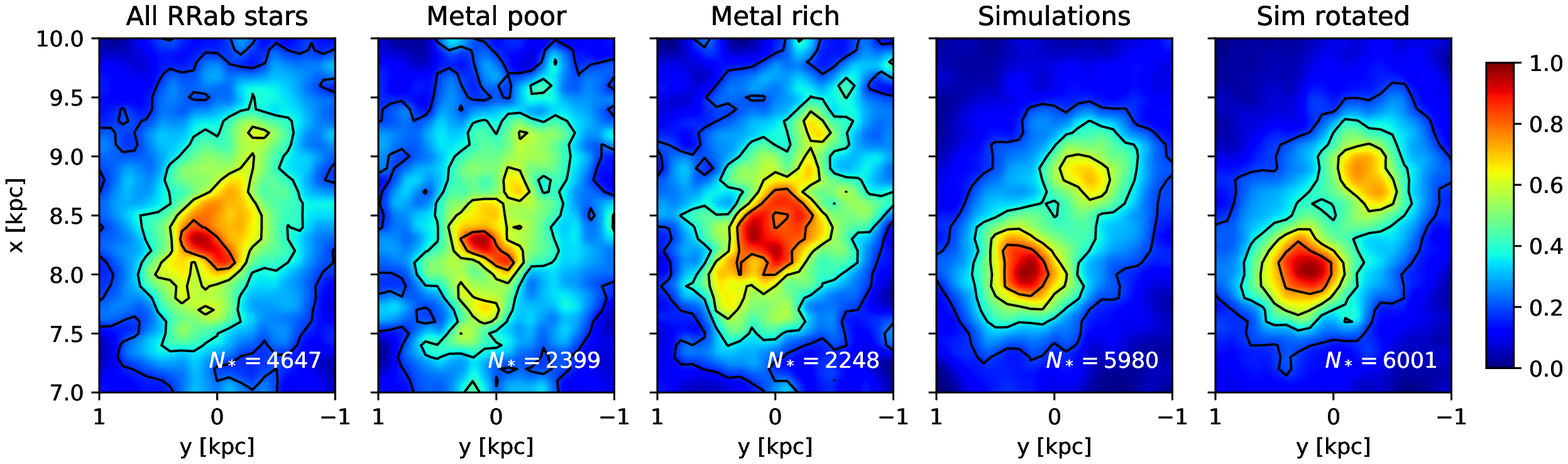}
    \caption{Distributions projected onto the Galactic equatorial plane of the same subsets as in Fig.~\ref{front}. The Sun is at the origin and the GC at $(R_0,0)$, where we assumed $R_0=8.47\,$kpc for the simulation to match the RR Lyrae data. All data sets were selected in $-6.5\degr<b<-2.8\degr$, which translates to an oblique cut $-0.11<z/s<-0.049$ in height $z$ depending on distance $s$. The maps were constructed by a kernel density estimator with smoothing length 0.15\,kpc. A double-hump structure is clearly visible for the metal poor RRab stars and the simulation data (albeit with larger separation), where it is generated by the bar peanut and the asymmetry between back and front owed entirely to the oblique cut. 
    \added{Except for minor differences, the first three panels of this figure are identical to Figs.~8 \& 9 of \protect\cite{Du2020}.}
    }
    \label{top}
\end{figure*}

We now investigate whether these predictions of kinematic fractionation hold for RR Lyrae stars in the Galactic bulge. According to stellar evolution theories \citep[e.g.][]{Marconi2015} and globular cluster studies \citep[e.g.][]{Walker1989, Clement2001}, RR Lyrae stars \added{are generally found in} populations older than $\sim10\,{\rm Gyr}$. However the precise age distribution and possible contamination by e.g. binary evolution are still debated \citep{Giacobbo2020}.

The top three panels of Fig.~\ref{front} display the distribution of our RRab stars in galactic coordinates; the bottom two panels show the oldest stars from our simulation, observed from the two possible Solar positions (bar angle 33\degr). The plots from the simulations were made by mirroring the particle distribution with respect to $b=0^{\circ}$ in order to increase the resolution. The density, in particular near the upper edge, shows two clear maxima both in the data and the simulation. For the latter, these are generated by the peanut structure presented in Section~\ref{sec:sims}. For the RRab stars, the two maxima are closer to each other than in the simulation and also extend to larger $|b|$. These quantitative differences indicate that the old MW bulge is radially smaller and vertically thicker than the simulation. \added{Note, however, that in our simulations we have not yet introduced GMCs, which will vertically heat the disc and thus influence the height difference between observations and simulation.}

\added{
Fig.~\ref{fig:minimum} shows star counts in galactic longitude from the latitude range where the peaks are most pronounced. The data show a clear bi-modality, or equivalently a minimum at $\ell\approx-0.6\degr$. We find the significance of this minimum to exceed $3\sigma$ (see figure caption for details).
The location of this minimum at negative longitudes is exactly where one expects it (for latitudes sufficiently away from the midplane $b=0$) from the geometry and orientation of the peanut. Similarly, the depth of the minimum increases with increasing distance from the midplane, which is also expected from geometry. These two facts make this result even more significant.
}

In Fig.~\ref{top} we add distance $s$ to the galactic coordinates to produce a face-on view of the data. 
Similarly to the Fig.\ref{front} the particle distribution from the simulations were mirrored with respect to the galactic plane. The cut of the sample at $b<-2.8\degr$ translates into a distance dependent cut in height, which crops the near peak less than the farther one. The degree to which the peanut humps appear connected in this projection depends both on the cut in latitude and on the profile of the bar peanut (as visualised in Fig.~7 of \citealt{Ness2012}). The simulation plots in this figure also demonstrate the degree to which noise affects these maps: there are various mostly small-scale features that appear in only one of these two equivalent maps and are hence indicative of the noise level expected also for the data.

In the leftmost panel of Fig.~\ref{top}, showing the full RRab sample, we see a similar geometrical effect with a clear first peak in front of the Galactic Centre (GC) and a smeared-out weaker tail in place of the second peak behind the GC. As already mentioned, the MW peanut structure appears to be radially smaller and vertically thicker than our simulation, and therefore the peaks are closer and appear more connected in this projection. For the metal poor stars (second panel of Fig.~\ref{top}), we can actually see the far peak (at $\sim8.8$\,kpc) separated from the first.

The fact that this separation is visible in the metal poor stars, but not in the metal rich stars, is consistent with kinematic fractionation, because the more metal-poor RR Lyrae stars the younger they are \citep{Solima2014}.
This implies that the metal-rich population either has too weak a peanut to be resolved with this number of stars or is not part of the peanut but some kind of ancient spheroid.

We have visualised our data and simulations in two different ways, once looking at the sky projection (Figs~\ref{front}\&~\ref{fig:minimum}), corresponding to an inclined near-end-on view of the bar, and once in a face-on view of the bar (Fig.~\ref{top}) constructed from the RR Lyrae distances. When using distances, the peanut signature was much less pronounced and only present for the metal poor subset of RRab stars. This is readily explained, since the separation between the density maxima is just as wide as the substantial distance uncertainties ($\sigma_{s} \approx 0.56\,{\rm kpc}$ on average, \citealt{Pietrukowicz2015}), to which one has to add uncertainties from the photometric metallicities \citep{Dekany2021}. 
Nevertheless, we added the face-on analysis in Fig.~\ref{top} for one important aspect, not covered by the sky projections of Fig~\ref{front}: the left peak is indeed the nearer one and the overall distribution is consistent with bar/peanut morphology. Beyond that, a more detailed analysis of the face-on distribution  require improvements on the distances.

%%%%%%%%%%%%%%%%%%%%%%%%%%%
\section{Discussion and conclusion}
\subsection{Observational context}
Several previous studies attempted to look for the peanut structure in the RR Lyrae stars and concluded that it is either absent or RR Lyrae stars generally do not trace the bar. \cite{Debattista2017} used the sample of \cite{Dekany2013} and both argued for a \added{spheroidal} distribution of the central RR Lyrae population. Their sample is slightly less populous (by $\sim10\%$). Mostly, it appears that previous studies were held back by using the mean distance of stars vs.\ longitude as a primary indicator. This is a sub-optimal choice because this indicator strongly depends on the density profiles/extent of the distribution perpendicular to the bar major axis, which pushes the mean distance towards a constant value - an effect that cannot be corrected if we do not have precise information on the density profiles \citep[see also][]{Pietrukowicz2015}. Also selection function and chosen boundary box further affect this measure.

\cite{Dekany2013} found \added {in fact} indication for a peanut-like substructure in \added{the spatial distribution (akin to Fig.~\ref{top})} below $b < -5 \degr$\added{with $2\sigma$ significance for bi-modality from a KS-like test. They disregard this in favour of an interpretation in terms of a spheroidal distribution}. However, their result matches with our findings at lower latitude and thus adds to the significance of finding a peanut shaped structure. 

\cite{Kunder2016, Kunder2020} analysed RR Lyrae stars with radial velocity information and argued that they are a part of the spherical bulge-like component. However, their data do not fully sample the central region ($-3\degr<l<3\degr$) and mostly the area where we find the peanut signatures. Morerover any weak signal in their sample may be hidden by selection effects.

Finally, \cite{Pietrukowicz2015} looked for signatures of the peanut in a sub-sample of the stars studied in this letter. They concluded the lack of such structure similar to that found in red clump stars by \cite{Ness2012}, but did not detect the 2-3 times smaller peanut we have found.

Recently, \cite{Grady2020} used a period-age relationship for Mira variables in Gaia DR2 \citep{Mowlavi2018} to analyse the age-structure of disc and bulge, indicating that the younger part of their sample shows a peanut-like appearance, with less discernible structure for the oldest stars. We re-examined those old stars in their sample and found that this is due to low number statistics precluding any conclusions.

\subsection{Theoretical implications}
The central finding of this work is evidence in both simulations and observations of a different distance between the density maxima of the bar peanut structure depending on stellar population. We point out that the distance between the density peaks for each single population increases towards higher altitude. Future analysis should disentangle this clearly from the difference in distance between density maxima for different populations at a fixed altitude (the peanut spread factor found in this work). 

The face-on maps on Fig.~\ref{top} suggest that the separation of the peanut humps in RR Lyrae stars is about 0.7\,kpc, which is also consistent with a projected distance of $\sim2.5\degr$ between the two peaks in the sky projection (Figs.~\ref{front}\&\ref{fig:minimum}, assuming a bar angle of $\sim30\degr$). Contrasting this with the value of $\sim3.3\,$kpc derived from near-IR images \citep{Ciambur2017} \added{or from red-clump stars \citep{Wegg2013}} sets an upper limit (due to unknown increase with altitude) to the peanut spread factor of 4-5 between the respective BP structures. In our simulation the peanut spread factor was reaching a value of 2-3 between the oldest populations and those born 4-5\,Gyr later. In our otherwise equivalent simulations without inside-out disc formation, this factor was significantly (at least $25 \%$) smaller.  We note for comparison that \cite{Buck2018} found a small peanut spread factor (due to their wide latitude range it is not clear how much of the difference might be spreading with altitude) from simulations with a cosmological setup.

Ultimately, anything that affects bar evolution, \added{such as angular-momentum transport from gas and to dark-matter, star formation (as function of space and time), minor mergers, etc.}, can also influence the peanut-spread factor. Presently, we have no detailed qualitative and quantitative knowledge on these effects. The upper limit for the peanut spread factor of 4-5, suggested by the sample comparison, appears larger than what our and other simulations are able to reproduce so far. If confirmed by more accurate data and larger samples, this particular finding calls for deeper theoretical studies. Independent of this peculiar finding, the peanut spread factor is an interesting constraint on the growth/formation of the bar and disc. \added{We anticipate that it will be possible to constrain the bar formation time by comparing the peanut spread of different populations with independent measures of the inside-out formation.}

\section*{Acknowledgements}
\added{We thank the annonymous reviewer for useful comments.}
This work was supported by the STFC grant ST/S000453/1. MS thanks B.-E.\ Semczuk for support. EA thanks the CNES for financial support. We appreciate insightful discussions with P. Pietrukowicz, I.\ Ebrova, J.\ Eliasek and \added{S.\ Vaughan}.  This work was performed using the DiRAC Data Intensive service at Leicester, operated by the University of Leicester IT Services, which forms part of the STFC DiRAC HPC Facility (\url{www.dirac.ac.uk}). The equipment was funded by BEIS capital funding via STFC capital grants ST/K000373/1 and ST/R002363/1 and STFC DiRAC Operations grant ST/R001014/1. DiRAC is part of the National e-Infrastructure.

%%%%%%%%%%%%%%%%%%%%%%%%%%%%%%%%%%%%%%%%%%%%%%%%%%
\section*{Data Availability}

The RR Lyrae stars data comes from the publicly available OGLE Collection of Variable Stars \url{https://ogledb.astrouw.edu.pl/~ogle/OCVS/}. The simulation data can be shared on reasonable request to the corresponding author.



\bibliographystyle{mnras}
\bibliography{example} % if your bibtex file is called example.bib

\begin{thebibliography}{}
\makeatletter
\relax
\def\mn@urlcharsother{\let\do\@makeother \do\$\do\&\do\#\do\^\do\_\do\%\do\~}
\def\mn@doi{\begingroup\mn@urlcharsother \@ifnextchar [ {\mn@doi@}
  {\mn@doi@[]}}
\def\mn@doi@[#1]#2{\def\@tempa{#1}\ifx\@tempa\@empty \href
  {http://dx.doi.org/#2} {doi:#2}\else \href {http://dx.doi.org/#2} {#1}\fi
  \endgroup}
\def\mn@eprint#1#2{\mn@eprint@#1:#2::\@nil}
\def\mn@eprint@arXiv#1{\href {http://arxiv.org/abs/#1} {{\tt arXiv:#1}}}
\def\mn@eprint@dblp#1{\href {http://dblp.uni-trier.de/rec/bibtex/#1.xml}
  {dblp:#1}}
\def\mn@eprint@#1:#2:#3:#4\@nil{\def\@tempa {#1}\def\@tempb {#2}\def\@tempc
  {#3}\ifx \@tempc \@empty \let \@tempc \@tempb \let \@tempb \@tempa \fi \ifx
  \@tempb \@empty \def\@tempb {arXiv}\fi \@ifundefined
  {mn@eprint@\@tempb}{\@tempb:\@tempc}{\expandafter \expandafter \csname
  mn@eprint@\@tempb\endcsname \expandafter{\@tempc}}}

\bibitem[\protect\citeauthoryear{{Athanassoula}}{{Athanassoula}}{2018}]{Athanassoula2018}
{Athanassoula} E.,  2018, in {Chiappini} C.,  {Minchev} I.,  {Starkenburg} E.,
   {Valentini} M.,  eds, ~ Vol. 334, Rediscovering Our Galaxy. pp 65--72
  (\mn@eprint {arXiv} {1801.07720}), \mn@doi{10.1017/S1743921317008778}

\bibitem[\protect\citeauthoryear{{Athanassoula}, {Laurikainen}, {Salo}  \&
  {Bosma}}{{Athanassoula} et~al.}{2015}]{Athanassoula2015}
{Athanassoula} E.,  {Laurikainen} E.,  {Salo} H.,   {Bosma} A.,  2015, \mn@doi
  [\mnras] {10.1093/mnras/stv2231}, \href
  {https://ui.adsabs.harvard.edu/abs/2015MNRAS.454.3843A} {454, 3843}

\bibitem[\protect\citeauthoryear{{Athanassoula}, {Rodionov}  \&
  {Prantzos}}{{Athanassoula} et~al.}{2017}]{Athanassoula2017}
{Athanassoula} E.,  {Rodionov} S.~A.,   {Prantzos} N.,  2017, \mn@doi [\mnras]
  {10.1093/mnrasl/slw255}, \href
  {https://ui.adsabs.harvard.edu/abs/2017MNRAS.467L..46A} {467, L46}

\bibitem[\protect\citeauthoryear{{Aumer} \& {Sch{\"o}nrich}}{{Aumer} \&
  {Sch{\"o}nrich}}{2015}]{AumerSchoenrich2015}
{Aumer} M.,  {Sch{\"o}nrich} R.,  2015, \mn@doi [\mnras]
  {10.1093/mnras/stv2252}, \href
  {https://ui.adsabs.harvard.edu/abs/2015MNRAS.454.3166A} {454, 3166}

\bibitem[\protect\citeauthoryear{{Babusiaux}}{{Babusiaux}}{2016}]{Babusiaux2016}
{Babusiaux} C.,  2016, \mn@doi [\pasa] {10.1017/pasa.2016.1}, \href
  {https://ui.adsabs.harvard.edu/abs/2016PASA...33...26B} {33, e026}

\bibitem[\protect\citeauthoryear{{Buck}, {Ness}, {Macci{\`o}}, {Obreja}  \&
  {Dutton}}{{Buck} et~al.}{2018}]{Buck2018}
{Buck} T.,  {Ness} M.~K.,  {Macci{\`o}} A.~V.,  {Obreja} A.,   {Dutton} A.~A.,
  2018, \mn@doi [\apj] {10.3847/1538-4357/aac890}, \href
  {https://ui.adsabs.harvard.edu/abs/2018ApJ...861...88B} {861, 88}

\bibitem[\protect\citeauthoryear{Bureau \& Freeman}{Bureau \&
  Freeman}{1999}]{Bureau1999}
Bureau M.,  Freeman K.~C.,  1999, \mn@doi [The Astronomical Journal]
  {10.1086/300922}, 118, 126

\bibitem[\protect\citeauthoryear{{Catelan}, {Pritzl}  \& {Smith}}{{Catelan}
  et~al.}{2004}]{Catelan2004}
{Catelan} M.,  {Pritzl} B.~J.,   {Smith} H.~A.,  2004, \mn@doi [\apjs]
  {10.1086/422916}, \href
  {https://ui.adsabs.harvard.edu/abs/2004ApJS..154..633C} {154, 633}

\bibitem[\protect\citeauthoryear{{Ciambur}, {Graham}  \&
  {Bland-Hawthorn}}{{Ciambur} et~al.}{2017}]{Ciambur2017}
{Ciambur} B.~C.,  {Graham} A.~W.,   {Bland-Hawthorn} J.,  2017, \mn@doi
  [\mnras] {10.1093/mnras/stx1823}, \href
  {https://ui.adsabs.harvard.edu/abs/2017MNRAS.471.3988C} {471, 3988}

\bibitem[\protect\citeauthoryear{{Clement} et~al.,}{{Clement}
  et~al.}{2001}]{Clement2001}
{Clement} C.~M.,  et~al., 2001, \mn@doi [\aj] {10.1086/323719}, \href
  {https://ui.adsabs.harvard.edu/abs/2001AJ....122.2587C} {122, 2587}

\bibitem[\protect\citeauthoryear{{Debattista}, {Ness}, {Gonzalez}, {Freeman},
  {Zoccali}  \& {Minniti}}{{Debattista} et~al.}{2017}]{Debattista2017}
{Debattista} V.~P.,  {Ness} M.,  {Gonzalez} O.~A.,  {Freeman} K.,  {Zoccali}
  M.,   {Minniti} D.,  2017, \mn@doi [\mnras] {10.1093/mnras/stx947}, \href
  {https://ui.adsabs.harvard.edu/abs/2017MNRAS.469.1587D} {469, 1587}

\bibitem[\protect\citeauthoryear{{Dehnen}}{{Dehnen}}{2000}]{Dehnen2000}
{Dehnen} W.,  2000, \mn@doi [\apjl] {10.1086/312724}, \href
  {https://ui.adsabs.harvard.edu/abs/2000ApJ...536L..39D} {536, L39}

\bibitem[\protect\citeauthoryear{{Dehnen} \& {McLaughlin}}{{Dehnen} \&
  {McLaughlin}}{2005}]{DehnenMcLaughlin2005}
{Dehnen} W.,  {McLaughlin} D.~E.,  2005, \mn@doi [\mnras]
  {10.1111/j.1365-2966.2005.09510.x}, \href
  {https://ui.adsabs.harvard.edu/abs/2005MNRAS.363.1057D} {363, 1057}

\bibitem[\protect\citeauthoryear{{D{\'e}k{\'a}ny}, {Minniti}, {Catelan},
  {Zoccali}, {Saito}, {Hempel}  \& {Gonzalez}}{{D{\'e}k{\'a}ny}
  et~al.}{2013}]{Dekany2013}
{D{\'e}k{\'a}ny} I.,  {Minniti} D.,  {Catelan} M.,  {Zoccali} M.,  {Saito}
  R.~K.,  {Hempel} M.,   {Gonzalez} O.~A.,  2013, \mn@doi [\apjl]
  {10.1088/2041-8205/776/2/L19}, \href
  {https://ui.adsabs.harvard.edu/abs/2013ApJ...776L..19D} {776, L19}

\bibitem[\protect\citeauthoryear{{D{\'e}k{\'a}ny}, {Grebel}  \&
  {Pojma{\'n}ski}}{{D{\'e}k{\'a}ny} et~al.}{2021}]{Dekany2021}
{D{\'e}k{\'a}ny} I.,  {Grebel} E.~K.,   {Pojma{\'n}ski} G.,  2021, arXiv
  e-prints, \href {https://ui.adsabs.harvard.edu/abs/2021arXiv210705983D} {p.
  arXiv:2107.05983}

\bibitem[\protect\citeauthoryear{{Du}, {Mao}, {Athanassoula}, {Shen}  \&
  {Pietrukowicz}}{{Du} et~al.}{2020}]{Du2020}
{Du} H.,  {Mao} S.,  {Athanassoula} E.,  {Shen} J.,   {Pietrukowicz} P.,  2020,
  \mn@doi [\mnras] {10.1093/mnras/staa2601}, \href
  {https://ui.adsabs.harvard.edu/abs/2020MNRAS.498.5629D} {498, 5629}

\bibitem[\protect\citeauthoryear{{Erwin} \& {Debattista}}{{Erwin} \&
  {Debattista}}{2017}]{Erwin2017}
{Erwin} P.,  {Debattista} V.~P.,  2017, \mn@doi [\mnras]
  {10.1093/mnras/stx620}, \href
  {https://ui.adsabs.harvard.edu/abs/2017MNRAS.468.2058E} {468, 2058}

\bibitem[\protect\citeauthoryear{{Fragkoudi}, {Di Matteo}, {Haywood},
  {G{\'o}mez}, {Combes}, {Katz}  \& {Semelin}}{{Fragkoudi}
  et~al.}{2017}]{Fragkoudi2017}
{Fragkoudi} F.,  {Di Matteo} P.,  {Haywood} M.,  {G{\'o}mez} A.,  {Combes} F.,
  {Katz} D.,   {Semelin} B.,  2017, \mn@doi [\aap]
  {10.1051/0004-6361/201630244}, \href
  {https://ui.adsabs.harvard.edu/abs/2017A&A...606A..47F} {606, A47}

\bibitem[\protect\citeauthoryear{{Fragkoudi} et~al.,}{{Fragkoudi}
  et~al.}{2020}]{Fragkoudi2020}
{Fragkoudi} F.,  et~al., 2020, \mn@doi [\mnras] {10.1093/mnras/staa1104}, \href
  {https://ui.adsabs.harvard.edu/abs/2020MNRAS.494.5936F} {494, 5936}

\bibitem[\protect\citeauthoryear{{Giacobbo} \& {Mapelli}}{{Giacobbo} \&
  {Mapelli}}{2020}]{Giacobbo2020}
{Giacobbo} N.,  {Mapelli} M.,  2020, \mn@doi [\apj] {10.3847/1538-4357/ab7335},
  \href {https://ui.adsabs.harvard.edu/abs/2020ApJ...891..141G} {891, 141}

\bibitem[\protect\citeauthoryear{{Gonzalez}, {Rejkuba}, {Zoccali}, {Valenti},
  {Minniti}, {Schultheis}, {Tobar}  \& {Chen}}{{Gonzalez}
  et~al.}{2012}]{Gonzalez2012}
{Gonzalez} O.~A.,  {Rejkuba} M.,  {Zoccali} M.,  {Valenti} E.,  {Minniti} D.,
  {Schultheis} M.,  {Tobar} R.,   {Chen} B.,  2012, \mn@doi [\aap]
  {10.1051/0004-6361/201219222}, \href
  {https://ui.adsabs.harvard.edu/abs/2012A&A...543A..13G} {543, A13}

\bibitem[\protect\citeauthoryear{{Grady}, {Belokurov}  \& {Evans}}{{Grady}
  et~al.}{2020}]{Grady2020}
{Grady} J.,  {Belokurov} V.,   {Evans} N.~W.,  2020, \mn@doi [\mnras]
  {10.1093/mnras/stz3617}, \href
  {https://ui.adsabs.harvard.edu/abs/2020MNRAS.492.3128G} {492, 3128}

\bibitem[\protect\citeauthoryear{{Kunder} et~al.,}{{Kunder}
  et~al.}{2016}]{Kunder2016}
{Kunder} A.,  et~al., 2016, \mn@doi [\apjl] {10.3847/2041-8205/821/2/L25},
  \href {https://ui.adsabs.harvard.edu/abs/2016ApJ...821L..25K} {821, L25}

\bibitem[\protect\citeauthoryear{{Kunder} et~al.,}{{Kunder}
  et~al.}{2020}]{Kunder2020}
{Kunder} A.,  et~al., 2020, \mn@doi [\aj] {10.3847/1538-3881/ab8d35}, \href
  {https://ui.adsabs.harvard.edu/abs/2020AJ....159..270K} {159, 270}

\bibitem[\protect\citeauthoryear{{Marconi} et~al.,}{{Marconi}
  et~al.}{2015}]{Marconi2015}
{Marconi} M.,  et~al., 2015, \mn@doi [\apj] {10.1088/0004-637X/808/1/50}, \href
  {https://ui.adsabs.harvard.edu/abs/2015ApJ...808...50M} {808, 50}

\bibitem[\protect\citeauthoryear{{McWilliam} \& {Zoccali}}{{McWilliam} \&
  {Zoccali}}{2010}]{McWilliamZoccali2010}
{McWilliam} A.,  {Zoccali} M.,  2010, \mn@doi [\apj]
  {10.1088/0004-637X/724/2/1491}, \href
  {https://ui.adsabs.harvard.edu/abs/2010ApJ...724.1491M} {724, 1491}

\bibitem[\protect\citeauthoryear{{Mowlavi} et~al.,}{{Mowlavi}
  et~al.}{2018}]{Mowlavi2018}
{Mowlavi} N.,  et~al., 2018, \mn@doi [\aap] {10.1051/0004-6361/201833366},
  \href {https://ui.adsabs.harvard.edu/abs/2018A&A...618A..58M} {618, A58}

\bibitem[\protect\citeauthoryear{{Nataf}, {Udalski}, {Gould}, {Fouqu{\'e}}  \&
  {Stanek}}{{Nataf} et~al.}{2010}]{Nataf2010}
{Nataf} D.~M.,  {Udalski} A.,  {Gould} A.,  {Fouqu{\'e}} P.,   {Stanek} K.~Z.,
  2010, \mn@doi [\apjl] {10.1088/2041-8205/721/1/L28}, \href
  {https://ui.adsabs.harvard.edu/abs/2010ApJ...721L..28N} {721, L28}

\bibitem[\protect\citeauthoryear{{Ness} et~al.,}{{Ness}
  et~al.}{2012}]{Ness2012}
{Ness} M.,  et~al., 2012, \mn@doi [\apj] {10.1088/0004-637X/756/1/22}, \href
  {https://ui.adsabs.harvard.edu/abs/2012ApJ...756...22N} {756, 22}

\bibitem[\protect\citeauthoryear{{Pietrukowicz} et~al.,}{{Pietrukowicz}
  et~al.}{2015}]{Pietrukowicz2015}
{Pietrukowicz} P.,  et~al., 2015, \mn@doi [\apj] {10.1088/0004-637X/811/2/113},
  \href {https://ui.adsabs.harvard.edu/abs/2015ApJ...811..113P} {811, 113}

\bibitem[\protect\citeauthoryear{{Rojas-Arriagada} et~al.,}{{Rojas-Arriagada}
  et~al.}{2014}]{Rojas-Arriagada2014}
{Rojas-Arriagada} A.,  et~al., 2014, \mn@doi [\aap]
  {10.1051/0004-6361/201424121}, \href
  {https://ui.adsabs.harvard.edu/abs/2014A&A...569A.103R} {569, A103}

\bibitem[\protect\citeauthoryear{{Saito}, {Zoccali}, {McWilliam}, {Minniti},
  {Gonzalez}  \& {Hill}}{{Saito} et~al.}{2011}]{SaitoEtAl2011}
{Saito} R.~K.,  {Zoccali} M.,  {McWilliam} A.,  {Minniti} D.,  {Gonzalez}
  O.~A.,   {Hill} V.,  2011, \mn@doi [\aj] {10.1088/0004-6256/142/3/76}, \href
  {https://ui.adsabs.harvard.edu/abs/2011AJ....142...76S} {142, 76}

\bibitem[\protect\citeauthoryear{{Sch{\"o}nrich} \& {McMillan}}{{Sch{\"o}nrich}
  \& {McMillan}}{2017}]{SchoenrichMcmillan2017}
{Sch{\"o}nrich} R.,  {McMillan} P.~J.,  2017, \mn@doi [\mnras]
  {10.1093/mnras/stx093}, \href
  {https://ui.adsabs.harvard.edu/abs/2017MNRAS.467.1154S} {467, 1154}

\bibitem[\protect\citeauthoryear{{Smolec}}{{Smolec}}{2005}]{Smolec2005}
{Smolec} R.,  2005, \actaa, \href
  {https://ui.adsabs.harvard.edu/abs/2005AcA....55...59S} {55, 59}

\bibitem[\protect\citeauthoryear{{Sollima}, {Cassisi}, {Fiorentino}  \&
  {Gratton}}{{Sollima} et~al.}{2014}]{Solima2014}
{Sollima} A.,  {Cassisi} S.,  {Fiorentino} G.,   {Gratton} R.~G.,  2014,
  \mn@doi [\mnras] {10.1093/mnras/stu1564}, \href
  {https://ui.adsabs.harvard.edu/abs/2014MNRAS.444.1862S} {444, 1862}

\bibitem[\protect\citeauthoryear{{Soszy{\'n}ski} et~al.,}{{Soszy{\'n}ski}
  et~al.}{2014}]{Soszynski2014}
{Soszy{\'n}ski} I.,  et~al., 2014, \actaa, \href
  {https://ui.adsabs.harvard.edu/abs/2014AcA....64..177S} {64, 177}

\bibitem[\protect\citeauthoryear{{Soszy{\'n}ski} et~al.,}{{Soszy{\'n}ski}
  et~al.}{2019}]{Soszynski2019}
{Soszy{\'n}ski} I.,  et~al., 2019, \mn@doi [\actaa]
  {10.32023/0001-5237/69.4.2}, \href
  {https://ui.adsabs.harvard.edu/abs/2019AcA....69..321S} {69, 321}

\bibitem[\protect\citeauthoryear{{Udalski}, {Szyma{\'n}ski}  \&
  {Szyma{\'n}ski}}{{Udalski} et~al.}{2015}]{Udalski2015}
{Udalski} A.,  {Szyma{\'n}ski} M.~K.,   {Szyma{\'n}ski} G.,  2015, \actaa,
  \href {https://ui.adsabs.harvard.edu/abs/2015AcA....65....1U} {65, 1}

\bibitem[\protect\citeauthoryear{{Uttenthaler}, {Schultheis}, {Nataf}, {Robin},
  {Lebzelter}  \& {Chen}}{{Uttenthaler} et~al.}{2012}]{Uttenthaler2012}
{Uttenthaler} S.,  {Schultheis} M.,  {Nataf} D.~M.,  {Robin} A.~C.,
  {Lebzelter} T.,   {Chen} B.,  2012, \mn@doi [\aap]
  {10.1051/0004-6361/201219055}, \href
  {https://ui.adsabs.harvard.edu/abs/2012A&A...546A..57U} {546, A57}

\bibitem[\protect\citeauthoryear{{Walker}}{{Walker}}{1989}]{Walker1989}
{Walker} A.~R.,  1989, \mn@doi [\pasp] {10.1086/132470}, \href
  {https://ui.adsabs.harvard.edu/abs/1989PASP..101..570W} {101, 570}

\bibitem[\protect\citeauthoryear{{Wegg} \& {Gerhard}}{{Wegg} \&
  {Gerhard}}{2013}]{Wegg2013}
{Wegg} C.,  {Gerhard} O.,  2013, \mn@doi [\mnras] {10.1093/mnras/stt1376},
  \href {https://ui.adsabs.harvard.edu/abs/2013MNRAS.435.1874W} {435, 1874}

\bibitem[\protect\citeauthoryear{{Weiland} et~al.,}{{Weiland}
  et~al.}{1994}]{Weiland1994}
{Weiland} J.~L.,  et~al., 1994, \mn@doi [\apjl] {10.1086/187315}, \href
  {https://ui.adsabs.harvard.edu/abs/1994ApJ...425L..81W} {425, L81}

\makeatother
\end{thebibliography}

%\begin{thebibliography}{99}
%\end{thebibliography}

%%%%%%%%%%%%%%%%%%%%%%%%%%%%%%%%%%%%%%%%%%%%%%%%%%

%%%%%%%%%%%%%%%%% APPENDICES %%%%%%%%%%%%%%%%%%%%%

% \appendix

% \section{Some extra material}

% If you want to present additional material which would interrupt the flow of the main paper,
% it can be placed in an Appendix which appears after the list of references.

%%%%%%%%%%%%%%%%%%%%%%%%%%%%%%%%%%%%%%%%%%%%%%%%%%

% Don't change these lines
\bsp	% typesetting comment
\label{lastpage}
\end{document}